\documentclass{aip-cp}
\usepackage{graphicx}
\usepackage[numbers]{natbib}
\usepackage{rotating}
\usepackage{notes2bib}
\begin{document}

\title{Applications for ultimate spatial resolution in LASER based $\mu$-ARPES:
A FeSe case study}

\author[aff1]{E. F. Schwier\corref{cor1}}
\corresp[cor1]{Corresponding author: schwier@hiroshima-u.ac.jp}

\author[aff2]{H. Takita}
\author[aff2]{W. Mansur} 
\author[aff1]{A. Ino}
\author[aff1,aff3,aff4]{M. Hoesch} 
\author[aff3]{M. D. Watson} 
\author[aff5,aff6]{A. A. Haghighirad}
\author[aff1]{K. Shimada}

\affil[aff1]{Hiroshima Synchrotron Radiation Center, Hiroshima University, 2-313 Kagamiyama, Higashi-Hiroshima 739-0046, Japan}

\affil[aff2]{Graduate School of Science, Hiroshima University, 1-3-1 Kagamiyama, Higashi-Hiroshima 739-8526, Japan}

\affil[aff3]{Diamond Light Source, Harwell Campus, Didcot OX11 0DE, United Kingdom}

\affil[aff4]{DESY Photon Science, Deutsches Elektronen-Synchrotron, D-22603 Hamburg, Germany}

\affil[aff5]{Clarendon Laboratory, Department of Physics, University of Oxford, Parks Road, Oxford OX1 3PU, United Kingdom}

\affil[aff6]{Institute for Solid State Physics, Karlsruhe Institute of Technology, 76021 Karlsruhe, Germany}

\maketitle

\begin{abstract}
Combining Angle resolved photoelectron spectroscopy (ARPES) and a
$\mu$-focused Laser, we have performed scanning ARPES microscopy
measurements of the domain population within the nematic phase of
FeSe single crystals. We are able to demonstrate a variation of the
domain population density on a scale of a few 10 $\mu$m while constraining
the upper limit of the single domain size to less than 5 $\mu m$.
This experiment serves as a demonstration of how combining the advantages
of high resolution Laser ARPES and an ultimate control over the spatial
dimension can improve investigations of materials by reducing the
cross contamination of spectral features of different domains.
\end{abstract}

\section{Introduction}

In this paper we present data from the material FeSe, which demonstrate the
full potential of laser based $\mu$-ARPES to probe with both high energy resolution, giving precision on the anisotropic electronic state, and high spatial resolution, giving information on the distribution of orthorhombic domains in the nematic state. Interest in the material arises due to its two separate phase transitions; $T_s$ at 90~K marks an unusual tetragonal-orthorhombic distortion, entering the so-called nematic phase characterized by in-plane electronic anisotropy, and at $T_{c}=8$~K the system displays unconventional superconductivity \cite{Boehmer2018review}. In other Fe-based superconductors, a similar structural transition, $T_s$, precedes the onset of magnetic ordering in certain systems \cite{Fernandes2014}, but FeSe is unique in that it does not show antiferromagnetic order at any temperature. The interest in this phase transition has
been fueled by the intriguing observations of strong electronic effects
occurring around $T_{s}$, with the observation of a divergent ``nematic
susceptibility" in samples under tensile strain~\cite{boehmer2015,Tanatar2017}, as 
well as the pronounced influence of nematic order on the two-fold symmetric superconducting gap structure \cite{Sprau2017}.

High quality
single crystals of FeSe grow in the tetragonal space group P4/nmm.
Typical samples have dimensions of $1\times1\times0.05$~mm$^{3}$.
The crystallographic phase transition
to an orthorhombic low temperature phase at $T_{S}=90$~K~\cite{McQueen2009} is subtle but sharp.
The reduced symmetry of the low temperature phase makes the $a$ and
$b$ axes inequivalent. 
Unless controlled by uniaxial strain, the sample will naturally form regions
with the $a$ axis aligned along either of two orthogonal directions
with equal probability, analogous  to domain formation in ferromagnets cooled in zero field. 
These regions have been found to form domains
by optical microscopy techniques such as optical polarized light imaging~\cite{Tanatar2017}
and photo-induced reflectivity~\cite{luo2017}, as well as scanning
tunneling microscopy~\cite{Kasahara2014,Sprau2017}. A relation to
the anisotropy of the electronic structure was first established in
Rb-intercalated samples by combined structural and electronic structure
microscopy studies~\cite{speller2014,speller2015}. The interplay
between structural distortion and electronic structure was found to
be crucial for understanding the enhanced superconductivity up to
$T_{c}\approx30$~K. In all samples of FeSe as well as Rb-intercalated
FeSe domains of varying size and shape have been found.

Photoemission spectroscopy, and in particular the momentum-resolved
ARPES has been the chosen technique in many studies of FeSe through
the tetragonal-to-orthorhombic phase transition at $T_{S}$~\cite{maletz2014,nakayama2014,watson2015,zhang2015,fanfarillo2016,watson2016}.
However, the large size of the beam spot of conventional ARPES systems leads to the simultaneous
observation of two copies of the momentum-resolved structures, thus
complicating the analysis. It was thus attempted early on to force
the samples into only one crystallographic orientation by application
of slight uniaxial stress during cooling (de-twinning). ARPES data
from such samples have given an unambiguous view of all electronic
bands and their momentum structure and Fermi surface~\cite{ming2011,Shimojima2014,watson2017}.
Residual intensity from domains of suppressed orientation is, however,
persistently visible in such measurements~\cite{Shimojima2014,watson2017b}, 
and the external strain on the samples can cause experimental issues, such as surface bending and cracking. 
This motivates a second approach, which we take here, to shrink the 
size of the beam to become comparable with the domain size. In the ideal case one could 
directly measure photoelectrons from one domain only, but in a more realistic scenario corresponding to current capabilities, local strains in the
sample can give regions dominated by a particular domain orientation, giving strong contrast in the ARPES dispersions
measured at different locations.   

\section{Result and Discussion}

The experiments were performed at the $\mu$-LaserARPES machine at
the Hiroshima Synchrotron Radiation Center (HiSOR). A detailed performance
evaluation of the system has been described in \cite{iwasawa2017cf}.
For the experiment a photon energy of $h\nu=6.199$ eV was used. If
not stated otherwise the photons were linearly polarized with
an angle of 60$^{\circ}$ against p-polarized light. The
angle was chosen by comparing the relative intensity of the four hole
pocket bands crossing the Fermi level around the $\overline{\Gamma}$-point
and subsequently optimizing the polarization to maximize the visibility
of all four bands.

\begin{figure}[h]
\centerline{\includegraphics[scale=0.4]{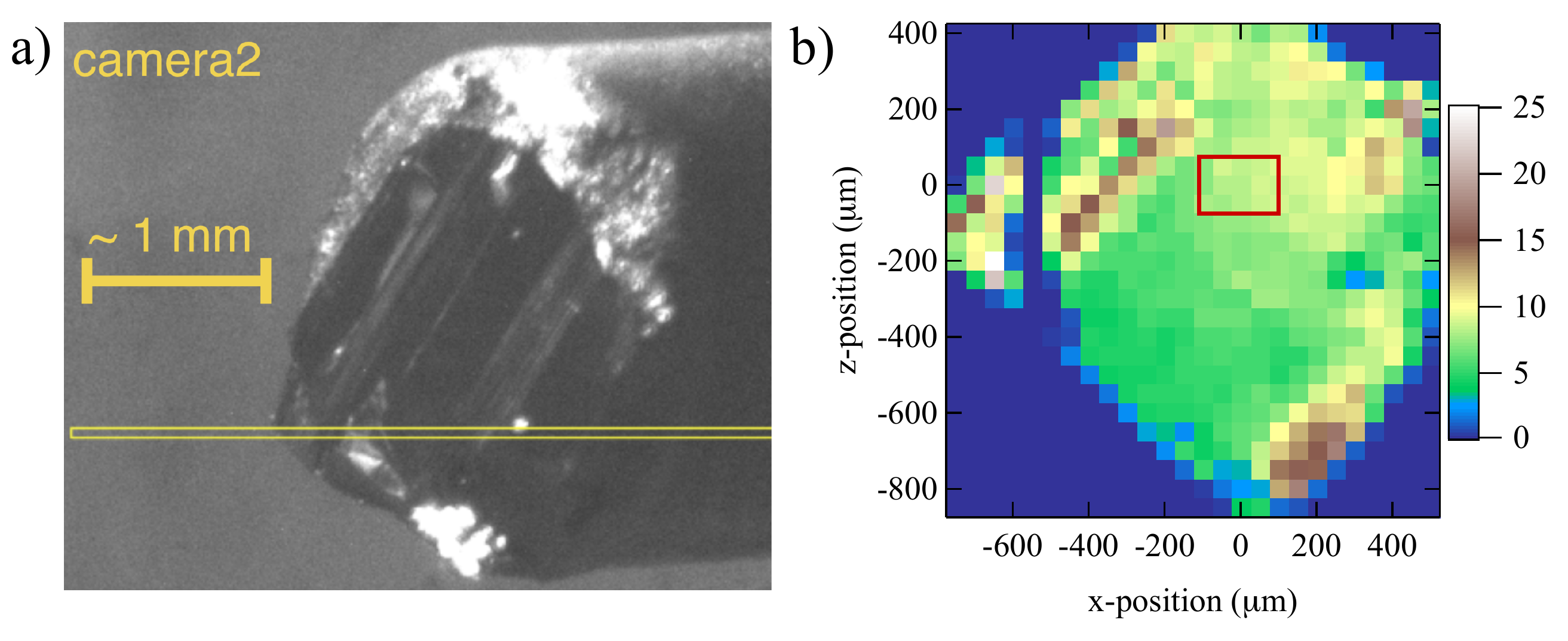}}
\caption{a) In-situ microscopic image of the cleaved FeSe surface. Different
cleaving surfaces are easily identified by the microscope. The yellow
rectangle is a guide to the eye corresponding to the incidence plane
of the Laser and the detection plane of the analyzer. b) Scanning
PES image of the sample topography. }
\label{realSpaceMicroscope}
\end{figure}

FeSe samples were grown by vapor transport \cite{watson2015} and
stored under Ar atmosphere before the measurement. Samples were cleaved
in-situ via the top-post method at a temperature of T = 20 K. During
the measurement the temperature was kept at 20 K via a resistive heater
mounted behind the sample holder inside the manipulator stage \cite{footnote1}.
Following the cleave, the homogeneity of the sample surface was first
verified via an optical long-distance microscope mounted to the measurement
chamber (Fig. \ref{realSpaceMicroscope} a). While the surface appears
not perfectly flat, past experience with FeSe tells that such a surface
can be considered a good cleave. From a LaserARPES point of view, care
has to be given to the absence of flakes and edge deformation
in order to reduce the risk of encountering unwanted distortion and
artefacts in the spectrum, a rather common
occurrence in low-energy ARPES measurements.

After obtaining an optically sufficient cleave, the homogeneity of
the surface was further verified by obtaining a rough spatial map
of the electronic structure (henceforth ``topography'') across the
sample. For this an ARPES measurement of the electronic band structure
at the Fermi level around the $\overline{\Gamma}$-point was performed
for each point within a 1.3$\times$ 1.3 mm area
using a step size of 50 $\mu m$. By integrating the total intensity
within the detector range Fig. \ref{realSpaceMicroscope} b) was obtained.
Since the data acquisition  allows for the manual analysis of the ARPES
spectrum at each point the choice of a suitable measurement area is
based on homogeneity as well as the experimenters judgement
of the ARPES features.

\begin{figure}[h]
\centerline{\includegraphics[width=0.9\columnwidth]{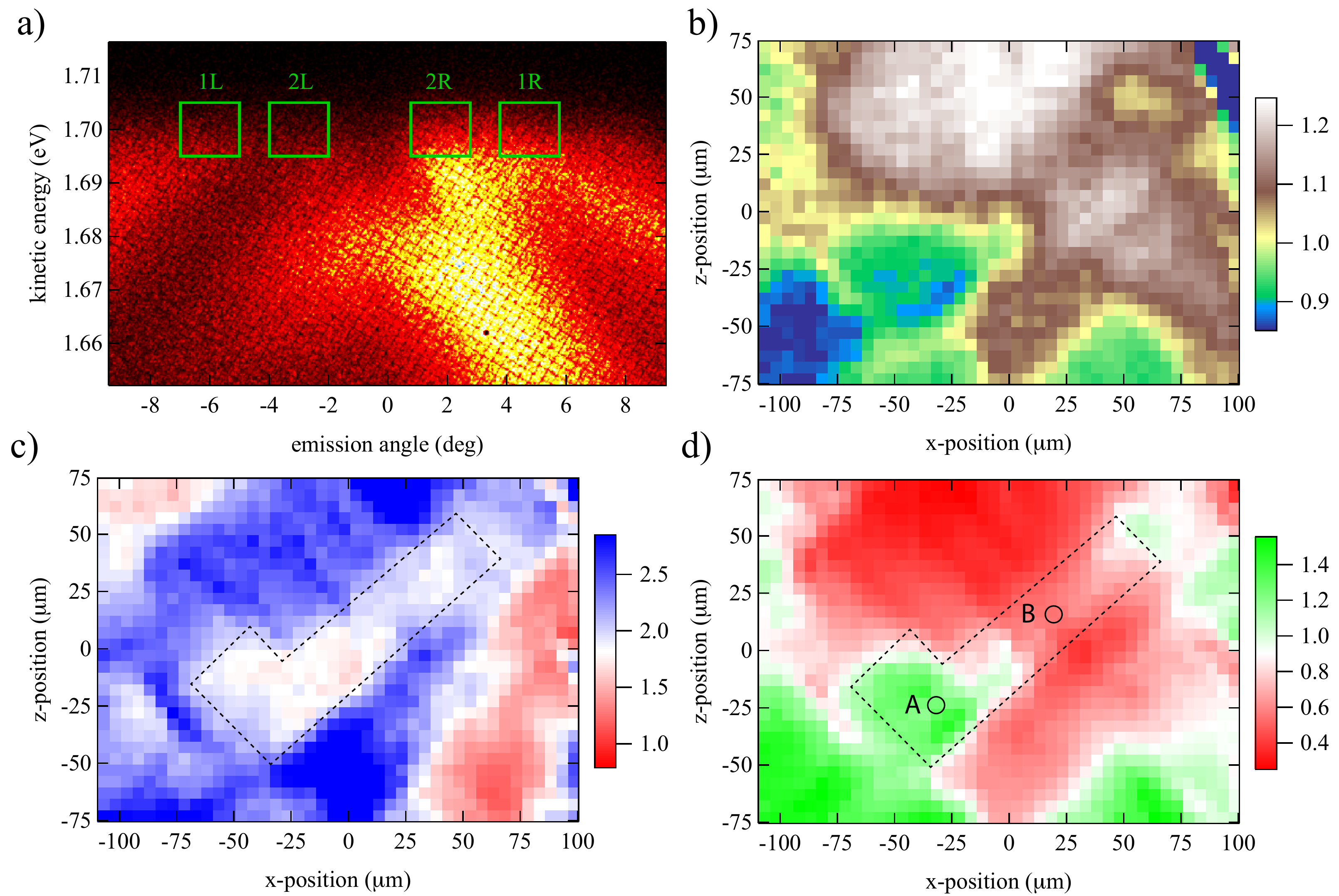}}
\caption{a) Example ARPES spectrum b) ``Topography'' of the sample obtained
by integrating the ARPES intensity across the detector. c) Plotting
R1/L1 as a figure of merit of constant orientation. Due to the strong
dependence of the spectral weight on the emission angle, a change
of the emission angle due to local electric fields or sample orientation
is easily detected as a deviation towards blue or red. d) Domain population
is calculated as a figure of merit by the ratio of R1/R2. Domain population
analysis is only carried out in the region previously identified as
having same orientation. The green and red areas correspond to majority
populations from the two domains of the nematic phase. Circles A and
B indicate the positions used for high resolution ARPES measurements
of the dispersion around the $\Gamma$-point.}
\label{zoomMaps}
\end{figure}

After choosing a suitable area (marked as the red rectangle in Fig
\ref{realSpaceMicroscope} b)) a second spatial ARPES map was obtained.
An area of 150$\times$ 210$ \mu m$ was chosen and
rastered with a step size of 5 $\mu m$. A representative example of a single
raw ARPES spectrum is shown in Fig. \ref{zoomMaps} a). Green rectangles
indicate the regions of interest (ROI) set up to analyze the spectral
weight from the inner and outer hole pocket bands. Each  are
expected to be related to a twinned orthorhombic domain in the nematic
phase of FeSe. Before moving to a detailed analysis of the obtained
data, the topography is again plotted as a reference of the
surface homogeneity in Fig. \ref{zoomMaps} b). Note that while clear
structures  are identifiable in the topography, the ARPES quality is of comparable quality 
across the measurement area. Furthermore, variation of the
total intensity has to be considered a sum effect 
of the real topography variations as well as matrix elements effects and sample orientation.

The goal of this study is to resolve the variation in population of
the two twinned orthorhombic domains, a task for which the topography
is not a suitable indicator. However, stored within the spatial
ARPES map are the spectral weight of the electronic structure at each point
in the mapping. Therefore, by a careful choice of ROIs corresponding
to distinct features in the ARPES spectra, one could process the data in a way
to obtain domain contrast. Care
has to be given to the fact that large amounts of data are automatically
processed. In such a case it is useful to define a quality parameter
first. 

Not only in the current case, but more generally, the sample orientation
can be used as a guide to detect the unwanted influence of intrinsic
(sample topography, surface bending) and extrinsic effects (local
electric fields, sample edge effects). If we consider the electronic
structure in each of the orthorhombic domains in FeSe we find
that the outer hole pocked (represented by ROIs: 1L \& 1R) and the
inner hole pocket (represented by ROIs: 2L \& 2R) have vastly different
dispersion parallel and perpendicular to the analyzer detection plane.
This fact leads to the choice of 1L/1R as a figure of merit representing
a constant orientation of the sample within our measurement range.
In Fig. \ref{zoomMaps} c) we have plotted this ratio. White color
represents an orientation corresponding to emission from a cut along
$\overline{\Gamma}$ (as judged  by manual review of the ARPES spectra), while red and blue represent areas with deviations
of the $k_{F}$. Based on this map, we choose to focus on the area
with constant orientation (dashed rectangle) for further analysis
of the domain population. While other parts of the sample will 
also exhibit different domain population, the loss of well-defined orientation
makes these areas unsuited for analysis under the current geometry
(manipulator polar and tilt angles).

Coming back to the original idea of attributing the two hole pockets
as representing a superposition of photoemission from the two twinned domains,
it is now required to chose how best to represent the contrast between
an overpopulation of each of the two domains. In the present case,
the ratio of 1R/2L preduces the highest contrast
as 1R and 2L exhibit the highest intensity variation across the selected
area of interest. Plotting this ratio in Fig. \ref{zoomMaps} d)
allowed us to identify regions on the surface with domain overpopulation.
It should be noted though that it is not possible to discuss absolute
domain population and therefore an intensity ratio of unity in Fig.
\ref{zoomMaps} d) does not necessarily correspond to a domain population
ratio of 1:1. Since the hole pockets are expected to originate from
different orbitals, it would be rather surprising if such a correspondence
could be found. Here the color scale was chosen in such a way
that within the rectangle of constant orientation, the
lowest and highest values would be represented as the brightest green
and red respectively. 

Based on the presence of spectral weight from both hole pockets in
all spectra of our map, the size of a single orthorhombic domain is
estimated to be well below the photon spot size of $\approx$5 $\mu m$.
To still be able to highlight the variation in domain population we
now choose two points (A and B) each representing a maximum and minimum
of the 1R/2L ratio for further detailed measurements. 

\begin{figure}[h]
\centerline{\includegraphics[width=0.9\columnwidth]{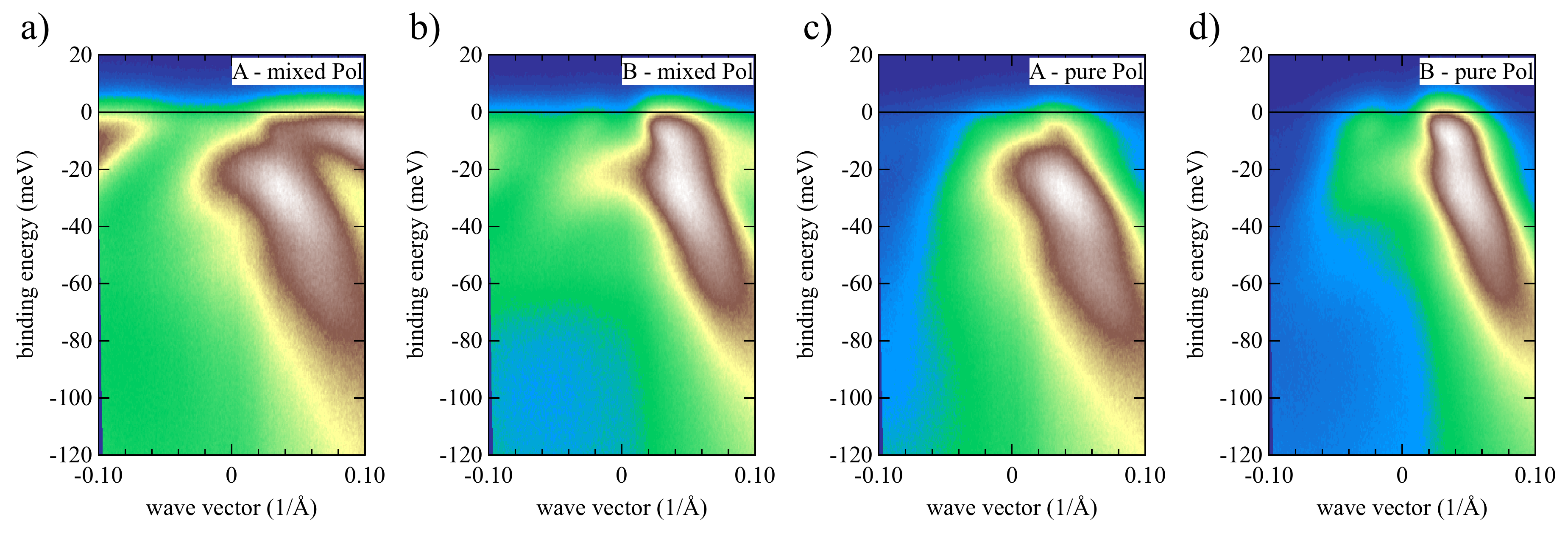}}
\caption{a) High symmetry cut at point A. The dominating intensity at the Fermi
level comes from the low dispersion hole pocket at larger $k_{F}$.
b) High symmetry cut at point B. The dominating intensity at the Fermi
level comes from the high dispersive hole pocket at smaller $k_{F}$.
c) High symmetry cut at point A using p-polarized light. The spectral
weight of the outer hole pocket is almost completely suppressed. d)
High symmetry cut at point B using p-polarized light showing similar
spectral weight as with mixed polarization in b).}
\label{HS_cuts}
\end{figure}

In Fig. \ref{HS_cuts} a) and b) we have plotted high resolution spectra
obtained from points A and B on the sample. It should be noted that
in both measurement points all four bands are crossing the Fermi level,
albeit with strongly differing intensities. While at A, the outer
hole pocket is dominating, at B we find the reverse situation. Since
the $k_{F}$ of the bands in both spectra are the same, the intensity
variation can only be explained by a variation of the domain population with the area
of sample being probed by the spot at the two locations.
Such a variation is in accordance with the assumption that twinned
domains coexist on the single crystalline surface within the nematic
phase. 

To illustrate that not only a high spatial resolution, but also the
choice of light polarization plays a significant role in resolving
the spatial variation of the spectral weight, we have plotted spectra
obtained with pure linear polarization in Fig. \ref{HS_cuts} c) and
d). Here the polarization was tuned to correspond to p-polarization
\cite{iwasawa2017cf} and both points, A and B were measured again.
While differences are still present between both spectra, they are
obviously less pronounced and moreover emission from the outer hole
pocket is strongly suppressed, making it much more difficult to determine
population differences employing pure linear polarization.

\begin{figure}[h]
\centerline{\includegraphics[width=0.9\columnwidth]{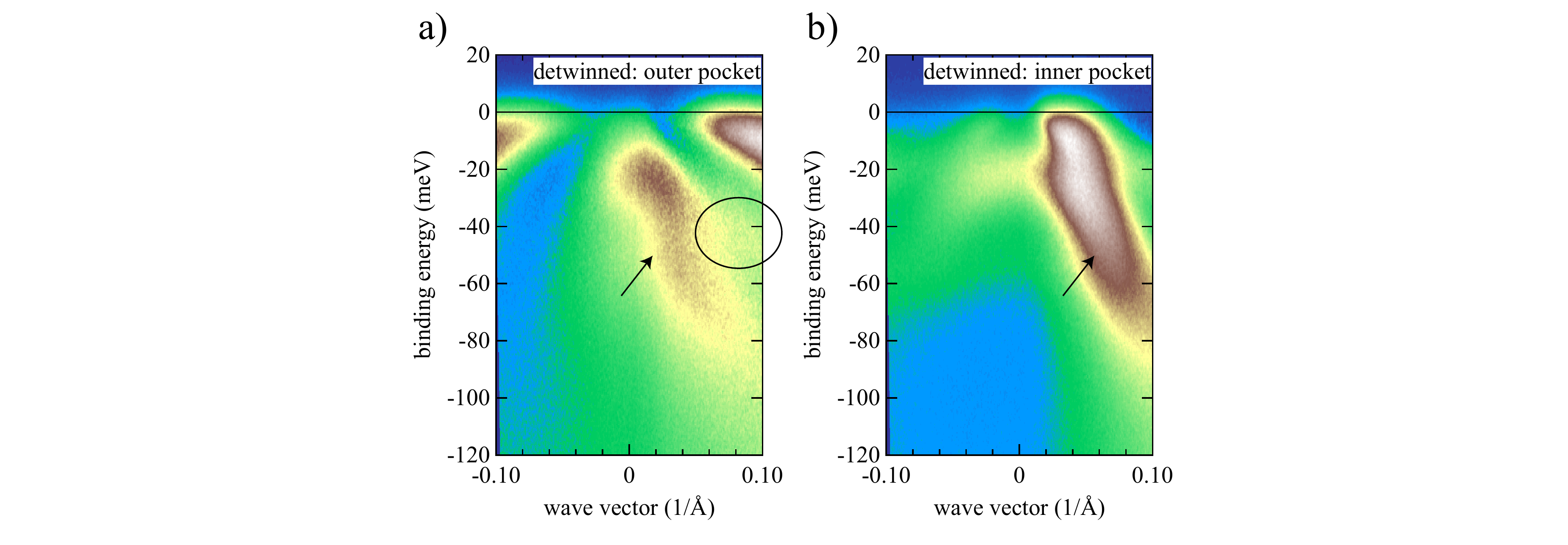}}
\caption{a) Numerical extraction of the emission of the outer hole pocket domain. Obtained by subtracting  emission from point B from emission at point A: outer = A - m $\times$ B.
b) Numerical extraction of the emission of the inner hole pocket domain. Obtained by subtracting  emission from point A from emission at point B: inner = B - n $\times$ A.}
\label{detwinned}
\end{figure}

Under the assumption that the spectra at both A and B are a linear
superposition of emission from the two twinned domains within the
area illuminated by the laser spot, it should be possible to separate
the emission from each domain by adjusting the relative weight ($n$
and $m$) of each spectra in Fig. \ref{HS_cuts} a) and b). The emission
from the domains with the outer and inner hole pockets ($outer=A-m\times B$
and $inner=B-n\times A$) can then be obtained by adjusting the normalization
factors in such a way that the minimum intensity in the resulting spectra does
not become negative, i.e. unphysical. The resulting spectra are shown
in Fig. \ref{detwinned} a) and b). At the Fermi level, the spectral
weight in each plot is entirely dominated by the outer and inner
hole pocket respectively, removing any visible cross-contamination
from the other domain. At higher binding energies the spectral weight
is less separated; the nematic effects are most pronounced near $E_F$, where
also the features are sharpest.  
The occupied and lighter inner hole pocket (arrows) is present in both spectra, 
though with unequal distribution of spectral weight, related via the matrix element 
effects to the different orbital character on the two cut directions. An additional 
heavier hole-like dispersion is detected in Fig. \ref{detwinned} a (circle).

Comparing these numerical
detwinning results with strain mediated detwinning \cite{watson2017}
as well as effective detwinning making use of light polarization \cite{Rhodes2018}
is however going beyond the scope of this paper, since the spectral
weight is highly dependent on the photoemission geometry and used
excitation energies. 

In conclusion, this work marks an important step towards demonstrating
the capabilities of laser-based $\mu$-ARPES. Choosing the well-studied
FeSe system allows for direct comparison of the advantages of surface
selectivity in our system compared to common Laser ARPES measurements
with spot sizes of the order of several 10 $\mu m$ and above as well
as modern high resolution ARPES synchrotron endstations with similar
photon spot size limitations. At the same time, this study could resolve
spatial inhomogeneity in the domain population of FeSe in unstrained
samples on the $\mu m$ scale opening up pathways to probe local strain
effects without the necessity to apply external strain. A follow-up
study should focus on the aspects of temperature dependence of the
spatial inhomogeneity as well as the shape and polarization dependence
of the Fermi surface with and without numerical detwinning. In the
absence of a deflection or image type analyzer, particularly the latter
task, while not impossible, poses a significant challenge to a $\mu$-ARPES
machine and puts a high demand on the sample surface quality due to
distortion of the local electric field. 
\section{Acknowledgments}
The authors would like to thank the Hiroshima N-Bard for supplying
liquid Helium. We also wish to thank H. Namatame and
M. Taniguchi for discussions and leadership that led to the creation
of the $\mu$-Laser ARPES machine. The high precision of the spatial
measurements are thanks to the ``nano-stage'' developed by Y. Aiura
and its software control developed by H. Iwasawa. Beamtime at
the HSRC was granted under proposal number 16BU008. A.A.H. acknowledges
the financial support of the Oxford Quantum Materials Platform Grant
(EP/M020517/1).

\nocite{*}
\bibliographystyle{aipnum-cp}%
\bibliography{FeSe_microscopy}%

\begin{thebibliography}{25}
\bibitem [{\citenamefont {{The use of the heater has been verified to not
  influence the ARPES measurement. No influence of a heating current induced
  magnetic field was found in the ARPES spectrum}}()}]{footnote1}%
  \BibitemOpen
  \bibfield  {author} {\bibinfo {author} {\bibnamefont {{The use of the heater
  has been verified to not influence the ARPES measurement. No influence of a
  heating current induced magnetic field was found in the ARPES spectrum}}},\
  }\href@noop {} {}\ \unskip\BibitemShut {NoStop}%
\bibitem [{\citenamefont {{Here we choose to set the color scale in such a way
  that within the chosen area of interest of constant orientation, the lowest
  and highest values would be represented as the brightest green and red
  respectively}}()}]{footnote2}%
  \BibitemOpen
  \bibfield  {author} {\bibinfo {author} {\bibnamefont {{Here we choose to set
  the color scale in such a way that within the chosen area of interest of
  constant orientation, the lowest and highest values would be represented as
  the brightest green and red respectively}}},\ }\href@noop {} {}\
  \unskip\BibitemShut {NoStop}%
\end{thebibliography}

\end{document}